\documentclass[superscriptaddress,twocolumn,10pt,prl,floatfix,nofootinbib,aps]{revtex4-2}

\input{glyphtounicode}
\usepackage[T1]{fontenc}
\usepackage[utf8]{inputenc}
\usepackage[english]{babel}
\usepackage{amsmath}  % needed for \tfrac, \bmatrix, etc.
\usepackage{amsfonts} % needed for bold Greek, Fraktur, and blackboard bold
\usepackage{amssymb}
\usepackage{graphicx} % needed for figures
\usepackage{hyperref}
\hypersetup{bookmarksopen,bookmarksnumbered,
            colorlinks,
            linkcolor=blue,
            citecolor=blue}

\usepackage{natbib}
\setcitestyle{super}
\usepackage{array}
\usepackage[table]{xcolor}
\usepackage{xcolor}
\usepackage{xr}
\usepackage{braket}
\usepackage{booktabs}
\usepackage{multirow}
\usepackage{dsfont} % for double lined 1 - identity symbol
\usepackage{caption}

\usepackage{tabularx}
\usepackage{siunitx}

\usepackage{bigstrut}
\usepackage{svg}
\usepackage{upgreek}
\usepackage{xfrac}
\usepackage{newfloat}

\DeclareFloatingEnvironment[name={\textbf{Supplementary Figure}}]{suppfigure}

\newcommand{\aref}[1]{\hyperref[#1]{Appendix~\ref*{#1}}}

\begin{document}

\captionsetup[figure]{name={FIG.},justification=justified,font=small,singlelinecheck=off}
\captionsetup{justification=justified,singlelinecheck=false}

\renewcommand{\equationautorefname}{Eq.}
\renewcommand{\figureautorefname}{Fig.}
\renewcommand*{\sectionautorefname}{Sec.}

\title{Coupling a \texorpdfstring{$^{73}$}{73-}Ge nuclear spin to an electrostatically defined quantum dot in silicon}

\author{Paul Steinacker}
\email{p.steinacker@unsw.edu.au}
\author{Gauri Goenka}
\author{Rocky Yue Su}
\affiliation{School of Electrical Engineering and Telecommunications, University of New South Wales, Sydney, NSW 2052, Australia}
\author{Tuomo Tanttu}
\author{Wee Han Lim}
\author{Santiago Serrano}
\author{Tim Botzem}
\author{Jesus D. Cifuentes}
\affiliation{School of Electrical Engineering and Telecommunications, University of New South Wales, Sydney, NSW 2052, Australia}
\affiliation{Diraq, Sydney, NSW, Australia}
\author{Shao Qi Lim}
\affiliation{School of Science, RMIT University, Melbourne, VIC, 3000, Australia}
\author{Jeffrey C. McCallum}
\affiliation{Centre for Quantum Computation and Communication Technology, School of Physics, The University of Melbourne, Parkville, Victoria, Australia}
\author{Brett C. Johnson}
\affiliation{School of Science, RMIT University, Melbourne, VIC, 3000, Australia}
\author{Fay E. Hudson}
\author{Kok Wai Chan}
\affiliation{School of Electrical Engineering and Telecommunications, University of New South Wales, Sydney, NSW 2052, Australia}
\affiliation{Diraq, Sydney, NSW, Australia}
\author{Christopher C. Escott}
\affiliation{Diraq, Sydney, NSW, Australia}
\author{Andre Saraiva}
\affiliation{Diraq, Sydney, NSW, Australia}
\author{Chih Hwan Yang}
\affiliation{School of Electrical Engineering and Telecommunications, University of New South Wales, Sydney, NSW 2052, Australia}
\affiliation{Diraq, Sydney, NSW, Australia}
\author{Vincent Mourik}
\affiliation{JARA Institute for Quantum Information (PGI-11), Forschungszentrum J\"ulich GmbH, 52425 J\"ulich, Germany}
\author{Andrea Morello}
\affiliation{School of Electrical Engineering and Telecommunications, University of New South Wales, Sydney, NSW 2052, Australia}
\author{Andrew S. Dzurak}
\affiliation{School of Electrical Engineering and Telecommunications, University of New South Wales, Sydney, NSW 2052, Australia}
\affiliation{Diraq, Sydney, NSW, Australia}
\author{Arne Laucht} 
\email{a.laucht@unsw.edu.au}
\affiliation{School of Electrical Engineering and Telecommunications, University of New South Wales, Sydney, NSW 2052, Australia}
\affiliation{Diraq, Sydney, NSW, Australia}

\date{\today}

\begin{abstract}
Single nuclear spins in silicon are a promising resource for quantum technologies due to their long coherence times and excellent control fidelities. Qubits and qudits have been encoded on donor nuclei, with successful demonstrations of Bell states and quantum memories on the spin-$\sfrac{1}{2}$ $^{31}$P and cat-qubits on the spin-$\sfrac{7}{2}$ $^{123}$Sb nuclei. Isoelectronic nuclear spins coupled to gate-defined quantum dots, such as the naturally occurring $^{29}$Si isotope, possess no additional charge and allow for the coupled electron to be shuttled without destroying the nuclear spin coherence. Here, we demonstrate the coupling and readout of a spin-$\sfrac{9}{2}$ $^{73}$Ge nuclear spin to a gate-defined quantum dot in SiMOS. The $^{73}$Ge nucleus was implanted by isotope-selective ion-implantation. We observe the hyperfine interaction (HFI) to the coupled quantum dot electron and are able to tune it from \SI{180}{\kilo \hertz} to \SI{350}{\kilo \hertz}, through the voltages applied to the lateral gate electrodes. This work lays the foundation for future spin control experiments on the spin-$\sfrac{9}{2}$ qudit as well as more advanced experiments such as entanglement distribution between distant nuclear spins or repeated weak measurements.
\boldmath
\textbf{}
\unboldmath
\end{abstract}
\maketitle
\noindent
Single nuclear spins in silicon are among the most coherent qubits, exhibiting exceptionally long coherence times~\cite{muhonen_storing_2014} and high 1-qubit control and 2-qubit entangling fidelities~\cite{madzik_precision_2022,thorvaldson_grovers_2025,zhang2025demonstrationquantumerrordetection}. 
As the magnetic moment of a single nucleus is very small, they are usually addressed through an accompanying, hyperfine-coupled electron~\cite{pla_high-fidelity_2013}. 
Group-V dopant atoms or crystal defects in silicon experience a pronounced hyperfine coupling with the nuclear spins of their host atoms ($\sim$\SI{100}{\mega \hertz})~\cite{neumann_single-shot_2010,pla_high-fidelity_2013} and surrounding lattice nuclei ($\sim$\SI{1}{}--\SI{8}{\mega \hertz})~\cite{childress_coherent_2006,pla_coherent_2014,laucht_high-fidelity_2014}. This arises from the highly confined electron wavefunction in such systems, which is bound by the Coulomb potential of the dopant's or defect's electric charge. 
In comparison, in gate-defined quantum dots, the electron wavefunction is typically less confined and overlaps with a much larger number of nuclear spins~\cite{slinker_quantum_2005,assali_hyperfine_2011,pla_coherent_2014,zhao2019single,lawrie_quantum_2020,hensen_silicon_2020}. 
This leads to a maximum hyperfine coupling of $\sim\SI{0.4}{}-\SI{1}{\mega \hertz}$ in silicon metal–oxide–semiconductor (SiMOS) quantum dots, where the combination of strong confinement at the Si/SiO$_{2}$ interface and the use of nanoscale gate electrodes confines the electron wavefunction to $\leq \SI{10}{\nano \meter}$ in diameter in lateral direction and a couple of nanometers in vertical direction~\cite{hensen_silicon_2020,witzel_remarkable_2022}. 
When using $^{28}$Si-isotopically enriched silicon host material with good electron spin coherence times and narrow electron spin resonance linewidths~ \cite{itoh_isotope_2014,muhonen_storing_2014,veldhorst2014addressable}, this hyperfine coupling is sufficient for addressing individual nuclear spins~\cite{hensen_silicon_2020,witzel_remarkable_2022}.

Most quantum error correction architectures require long-distance coupling of remote qubit sites~\cite{pica_surface_2016,veldhorst_silicon_2017,li_crossbar_2018,morello_donor_2020,gonzalez-zalba_scaling_2021,kunne_spinbus_2024,siegel_towards_2024}. The most promising approach in gate-defined quantum dots is shuttling~\cite{yoneda_coherent_2021,noiri_shuttling-based_2022,seidler_conveyor-mode_2022,zwerver_shuttling_2023,xue_sisige_2024,de_smet_high-fidelity_2025,lin_interplay_2025}, where electrons are physically moved along the interface while maintaining quantum information. 
In order to use shuttling in a quantum computing architecture with nuclear spins, the electron needs to be coupled and decoupled from the nuclear spin in a coherent manner. 
This requires deterministic movement of the coupled electron at a timescale that is much faster than the coupling, which is particularly challenging for strongly hyperfine-coupled donor electrons~\cite{hill2015surface,stemp_tomography_2024,stemp_scalable_2025}. Isoelectronic group-IV atoms in quantum dots offer an elegant alternative. The hyperfine interaction (HFI) is weaker than for donors and, as the electrostatic potential is defined using tuneable metallic gates, the electron can be swiftly moved away~\cite{yoneda_coherent_2021,noiri_shuttling-based_2022,seidler_conveyor-mode_2022,zwerver_shuttling_2023,xue_sisige_2024,de_smet_high-fidelity_2025,lin_interplay_2025}.

\begin{figure*}
    \centering
    \includegraphics[width = 1\textwidth]{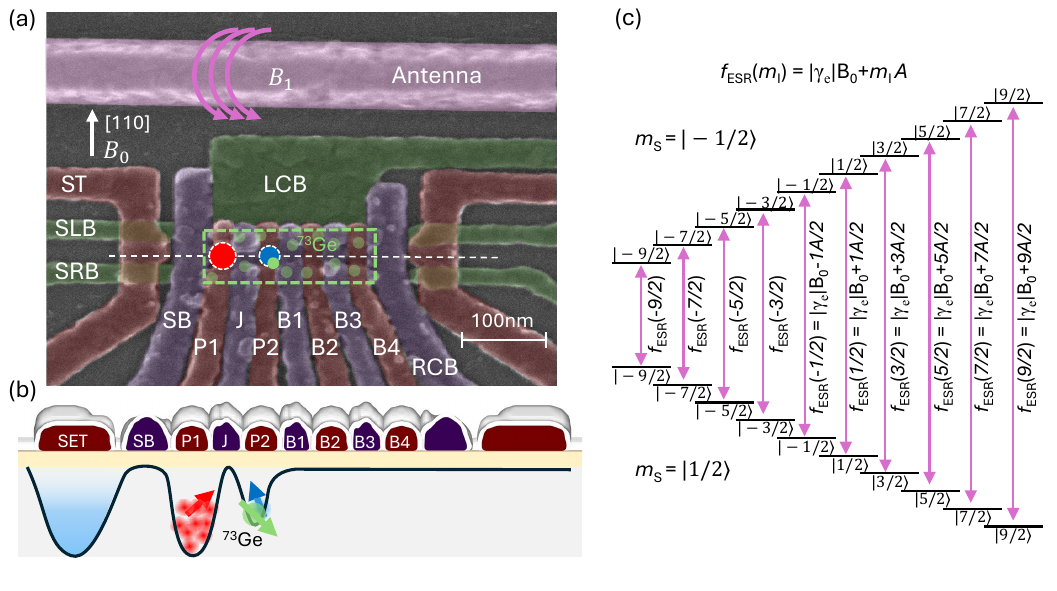}
    \caption{Device and electron-nucleus coupled system. 
    (a) Scanning electron micrograph of a nominally identical device. The four gate deposition layers are highlighted in green, red, purple, and pink, respectively. Arrows indicate the directions of the external applied d.c.\,magnetic field $B_{0}$ and the a.c.\,magnetic field $B_{1}$ the antenna generates. The formed quantum dots under the plunger gates P1 and P2 are indicated by the blue and red circle, respectively. The green circles indicate a possible isotopically-selected $^{73}$Ge nuclei distribution in the implantation window depicted by the green dashed box. The device is operated at the dilution refrigerator base temperature of $T=\SI{50}{\milli \kelvin}$. 
    (b) Schematic of the device's cross-section in the active region, marked as the dashed white line in (a). We form the two-dimensional electron gas at the Si/SiOx interface. The single-electron transistor is confined with the SLB, SRB, and SB. The quantum dots to form qubits are laterally confined with the LCB and RCB. The single-electron dot under P1 reveals a strongly coupled $I=9/2$ nuclear spin.
    (c) Energy level diagram of an isoelectronic $^{73}$Ge nucleus ($I=9/2$) coupled to an electron spin ($S=1/2$) of a quantum dot. The Zeeman interaction splits the states into an electron spin-up ($m_{\mathrm{S}}=1/2$) and electron spin down ($m_{\mathrm{S}}=-1/2$) manifold, which is then further split by the hyperfine interaction based on the nuclear spin quantum number $m_{\mathrm{I}} \in \{-9/2,-7,2,...,7/2,9/2\}$ and the hyperfine interaction strength $A$. In this coupled system, the electron spin resonance (ESR) frequency $f_{\mathrm{ESR}}$ depends on the nuclear spin orientation. All $2I+1=10$ different ESR transitions are equally spaced by $A$.  
}
    \label{fig:main_fig_1}
\end{figure*}

In this work, we demonstrate the coupling of a single spin-$\sfrac{9}{2}$ $^{73}$Ge nuclear spin to the electron spin of a SiMOS gate-defined quantum dot. We implanted the $^{73}$Ge isotope with a target concentration equivalent to \SI{800}{ppm} just below the Si/SiO$_{2}$ interface. Furthermore, the HFI is tunable between \SI{180}{}--\SI{350}{\kilo \hertz} by using the gate electrodes to change the electron wavefunction confinement and location. Our work lays the foundation for future spin control experiments on the spin-$\sfrac{9}{2}$ qudit, pointing toward the broader potential of isoelectronic species for scalable nuclear-spin-based quantum computing. Advanced initialization and control of high-dimensional nuclear spin systems have already been established for the $^{123}$Sb donor in silicon~\cite{asaad_coherent_2020,fernandez_de_fuentes_navigating_2024,yu_schrodinger_2025,vaartjes_certifying_2025}. Here, the bound electron has a larger HFI on the order of $A=$\SI{100}{}--\SI{200}{\mega \hertz}~\cite{morello_donor_2020}, which enables faster and more reliable nuclear spin readout. However, the smaller HFI of group IV nuclei has the advantage of maintaining coherence during electron movement~\cite{tosi_silicon_2017,hensen_silicon_2020}.

\section{Device fabrication and operation}
\begin{figure*}
    \centering
    \includegraphics[width = 1\textwidth]{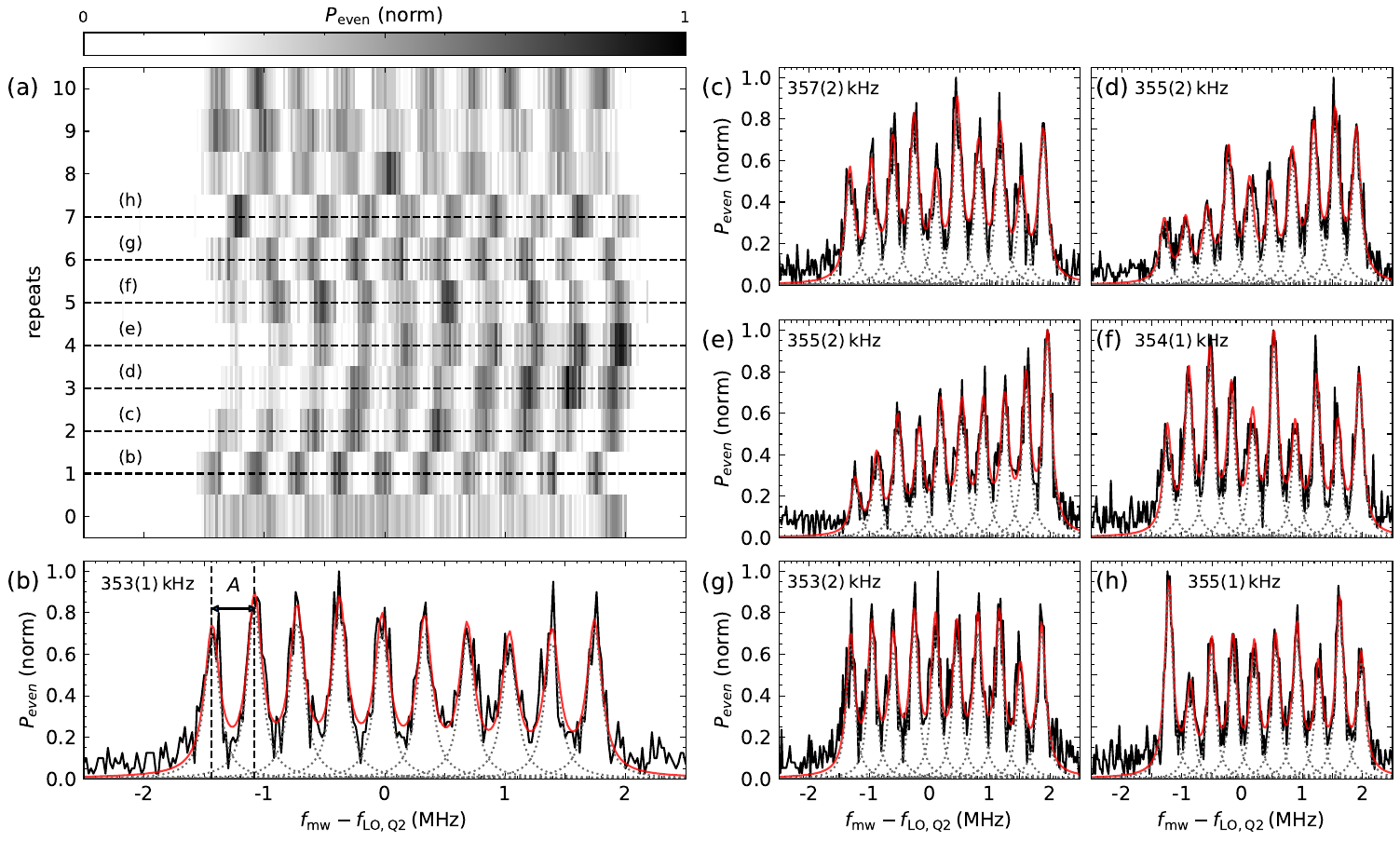}
    \caption{Nuclear spin signature in ESR. 
    (a) Normalized even electron spin probability $P_{\mathrm{even}}$ for repeated electron spin resonance probing as a function of applied microwave signal frequency. Each frequency step is averaged over 500 shots. 
    (b) Line-cut of the normalized electron spin-up probability $P_{\mathrm{even}}$ at repetition \#1 at the dashed black line in (a). The sum of 10 Lorentzian peaks (red line) fitted with equal spacing $A$. The peak spacing is the hyperfine interaction strength $A$ between the electron and nuclear spin. The dotted lines are the individual Lorentzian peak fits. The local oscillator frequency of qubit~2 is $f_{\mathrm{LO,Q2}} = \SI{8.401}{\giga \hertz}$.
    (c)-(h) Line-cut of the normalized electron spin-up probability $P_{\mathrm{even}}$ at repetitions \#2--7 at the dashed black lines in (a). Hyperfine interaction ranges from A = \SI{353 \pm 2}{}--\SI{357 \pm 2}{\kilo \hertz} across all repeats. Error bars represent a \SI{95}{\percent} confidence interval determined from the estimated covariance matrix.
}
    \label{fig:main_fig_2}
\end{figure*}

\noindent 
The quantum dots are electrostatically confined by a multi-layer aluminum gate-stack~\cite{angus_gate-defined_2007} and can be fabricated on top of an isotopically enriched ${}^{28}$Si substrate with $\SI{800}{ppm}$ residual $^{29}$Si~\cite{itoh_isotope_2014} and the equivalent of $\SI{800}{ppm}$ implanted ${}^{73}$Ge in the active region [see Figs.~\ref{fig:main_fig_1}(a,b)]. The implantation dose of \SI{800}{ppm} ${}^{73}$Ge was chosen because in previous work in Ref.~\cite{hensen_silicon_2020} on a similar device with residual \SI{800}{ppm} $^{29}$Si, two $I=1/2$ nuclear spins were hyperfine-coupled to the electron spin of a quantum dot. The device is biased such that the quantum dots are separated by approximately $\SI{60}{\nano\meter}$ occupying around $\SI{80}{\nano\meter^2}$ underneath the plunger gates (P1, P2) at the Si/SiO$_2$ interface. In the experiments, we operated the device in a configuration, where a double quantum dot with a charge configuration of (P1, P2) = (9,1), was kept isolated from the charge reservoir. Under the influence of an external d.c.\;magnetic field $B_0=\SI{0.3}{\tesla}$, the unpaired single electron spin in each quantum dot forms an effective two-level system utilized as a qubit~\cite{veldhorst_spin-orbit_2015,leon_coherent_2020}.

For single-shot charge readout, we used a radiofrequency single-electron transistor (RF-SET)~\cite{angus_silicon_2008} and integrated the signal at $\SI{205}{\mega \hertz}$ for $t_\mathrm{RO} = \SI{100}{\micro \second}$. We delivered the a.c.\;magnetic field $B_1$ that drives the electron spin state transitions via an on-chip antenna.

We initialized an odd electron spin state $\ket{\downarrow\uparrow}$ with a $t_\mathrm{init} = \SI{10}{\micro \second}$ ramp at $V_\mathrm{J} = \SI{3.2}{\volt}$ across the (8,2) to (9,1) inter-dot charge transition. We read out the spin state parity based on Pauli spin blockade \cite{seedhouse_pauli_2021}. Charge movement near the inter-dot charge transition from (9,1) to (8,2) is blockaded when both unpaired spins are parallel.

\section{Nuclear spin signature}
\noindent
We describe the joint electron-nuclear spin system under an external magnetic field $B_0$ along the z-direction by the simplified Hamiltonian 
\begin{align}
    H & = H_{\mathrm{Z,e}} + H_{\mathrm{Z,n}} + H_{\mathrm{HF}} \label{eq:Hamiltonian} \\
      & = B_0(\gamma_{\mathrm{e}}S_{\mathrm{z}} + \gamma_{\mathrm{Ge}}I_{\mathrm{z}}) + A(\mathbf{S} \cdot \mathbf{I}), 
\end{align}
with the electron spin operator $\mathbf{S}$ and nuclear spin operator $\mathbf{I}$. In Eq.~\ref{eq:HyperfineInteraction}, we assume that the general hyperfine tensor $\mathbf{A}$ is dominated by the isotropic component
\begin{align}
    A \;=\; \frac{2}{3}\,\hbar^{2}\,\mu_0\,\gamma_{\mathrm{e}}\,\gamma_{\mathrm{Ge}}\,|\Psi(0)|^2, \label{eq:HyperfineInteraction}
\end{align}
referred to as the Fermi-contact interaction, due to the quadratic dependence of the electron wavefunction overlap with the nucleus $|\Psi(0)|$. $\hbar$ is the reduced Planck's constant, $\mu_0$ the magnetic permeability in vacuum, $\gamma_\mathrm{e} = \SI{28.02}{\giga \hertz / \tesla}$ the gyromagnetic ratio of the electron, and $\gamma_{\mathrm{Ge}} = \SI{1.49}{\mega \hertz / \tesla}$ the gyromagnetic ratio of the $^{73}$Ge nuclear spin. In the experiments, we tuned the device electrically such that the HFI is maximized, which results in the anisotropic terms becoming negligible~\cite{van_de_walle_first-principles_1993,assali_hyperfine_2011,witzel_remarkable_2022}. The $^{73}$Ge isotope has a nuclear spin number of $I=9/2$, which gives rise to $2I+1 = 10$ possible nuclear spin states. The ten energy levels are separated by the nuclear Zeeman interaction and shifted by the HFI depending on the nuclear spin orientation [compare Fig.~\ref{fig:main_fig_1}(c)]. 

In the experiments, we resonantly drove the electron spin conditional on the nuclear spin state at a frequency of $f_{\mathrm{ESR}} = \left|\gamma_{\mathrm{e}}\right|B_{0}+m_\mathrm{I}A$, with $m_\mathrm{I} = $ -9/2, -7/2, ... , 7/2, 9/2. Figure~\ref{fig:main_fig_2}(a) shows the Larmor frequency as a function of measurement time or repeats, as we changed the applied microwave signal frequency $f_{\mathrm{mw}}$. In Fig.~\ref{fig:main_fig_2}(b), we fitted 10 Lorentzian peaks to repetition $\#1$ in Fig.~\ref{fig:main_fig_2}(a) and identified 10 peaks with equal spacing of the hyperfine constant $A=\SI{353 \pm 1}{\kilo \hertz}$, in agreement with the model introduced in Eq.~\ref{eq:Hamiltonian}--\ref{eq:HyperfineInteraction}.    

Furthermore, Fig.~\ref{fig:main_fig_2}(a) shows a small overall background frequency shift without a significant change of the HFI as a function of time/repeats, as evidenced by Fig.~\ref{fig:main_fig_2}(c)-(h). Again, we fitted 10 Lorentzian peaks to repetitions 2--7 in Fig.~\ref{fig:main_fig_2}(a) and extract HFI = $\SI{352 \pm 2}{} - \SI{357 \pm 2}{\kilo \hertz}$.

\begin{figure}
    \centering
    \includegraphics[width = 1\linewidth]{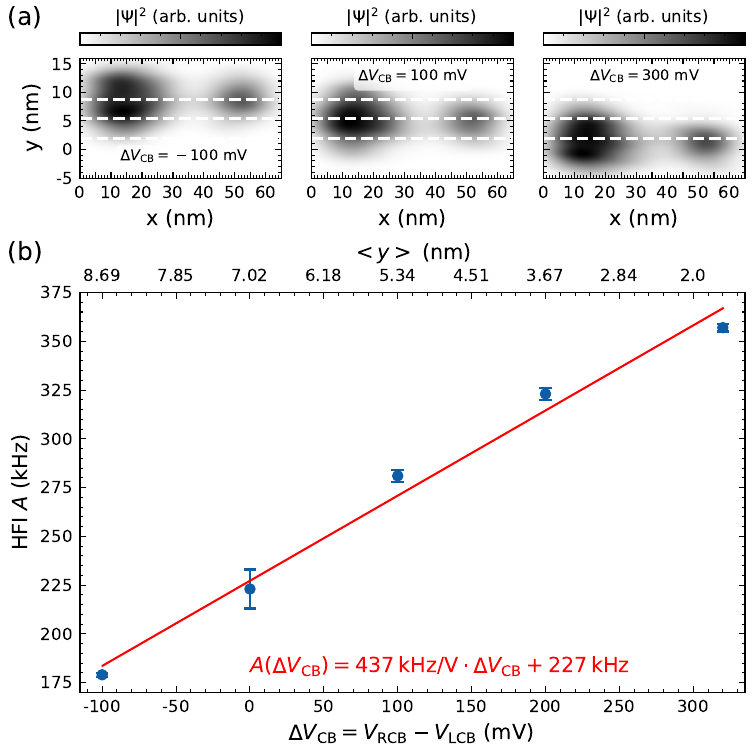}
    \caption{Hyperfine interaction tuning. 
    (a) Simulated electron wavefunction of the double-dot system for three different confinement voltages $\Delta V_{\mathrm{CB}}=\SI{-100}{\milli \volt}, \SI{100}{\milli \volt}, \SI{300}{\milli \volt}$. The white dashed lines indicate the respective wavefunction centre.
    (b) Hyperfine interaction strength $A$ of the quantum-dot-coupled $^{73}$Ge nuclear spin system as a function of confinement voltage $V_{\mathrm{CB}}$ (lower x-axis) and simulated mean of the electron density $\braket{y}$ (upper x-axis). The confinement voltage $V_{\mathrm{CB}}$ is the difference of the right and left confinement barrier voltage. Qualitatively, this corresponds to moving the center of the electron wavefunction closer to the $^{73}$Ge nuclear spin, leading to a larger Fermi-contact hyperfine interaction component. The red line is a linear fit with slope $ m=\SI{473}{\kilo \hertz / \volt }$ and vertical intercept of $A = \SI{227}{\kilo \hertz}$. Error bars represent a \SI{95}{\percent} confidence interval when determining the hyperfine constant from Larmor frequency time traces. 
}
    \label{fig:main_fig_3}
\end{figure}

\noindent
The hyperfine interaction is also tunable with the applied gate voltages. We increased the confinement voltage from $V_{\mathrm{CB}}=V_{\mathrm{RCB}}-V_{\mathrm{LCB}}=\SI{-0.25}{\volt}$ to $V_{\mathrm{CB}} = \SI{0.80}{\volt}$, which resulted in an estimated lateral shift of $\sim $\SI{7}{\nano \meter} and stronger confinement of the electron wavefunction close to the $^{73}$Ge nucleus [compare Fig.~\ref{fig:main_fig_3}(a)]. Figure~\ref{fig:main_fig_3}(b) shows the simulated shift of the electron density center and the thereby achieved tunability of the HFI ranging from $A=\SI{179 \pm 1}{\kilo \hertz}$ to the maximum value of $A=\SI{357 \pm 2}{\kilo \hertz}$, roughly doubling the coupling. 

Limitations in the experiment were a weakly effective J-gate, which lead to a lack of tunnel coupling and exchange interaction tunability. This reduced the observed electron spin readout visibility and inhibited deterministic electron spin initialization. Thus, we were not able to initialize the spin system by electron-nuclear-double resonance (ENDOR). 
Additionally, this limited the movement of 
the wavefunction mostly via the confinement barriers. With increased electrostatic control we expect to achieve even larger HFI.

\section{Conclusion}
\noindent
In conclusion, we measured the coupling of a single isoelectronic $^{73}$Ge isotope nuclear spin to the electron spin of an electrostatically defined quantum dot in silicon. Using the surrounding metal gates, we changed the electrostatic environment of the electron wavefunction to achieve tunability of the interaction strength by a factor of two. This opens a promising alternative approach to qudits in silicon reducing the constraints on maintaining phase coherence in donors. 
Having demonstrated access to the 10-dimensional nuclear spin Hilbert space in the $^{73}$Ge isotope lays the foundation for more sophisticated experiments in the future. Deterministic nuclear spin initialization, coherent nuclear control, and readout will enable a thorough characterization of the spin properties. In combination with the ability to coherently shuttle the electron to neighboring dots, this will enable more advanced experiments such as entanglement distribution between distant nuclear spins or repeated weak measurement of the nuclear spin. Ultimately, a single $^{73}$Ge physical qudit can be envisioned as a logical qubit utilizing a quantum error-correction encoding like moment angular system (MAUS)~\cite{gross_designing_2021,gross_hardware-efficient_2024}.

\bibliography{mybib}

\section*{Appendix}
\setcounter{section}{0}

\section{Experimental device}\label{methods:experimental_device}
\noindent
The device studied in this work was fabricated using multi-layer aluminum (Al) gate-stack silicon MOS technology on a isotopically enriched silicon-28 substrate with \SI{800}{ppm} residual $^{29}$Si. We implanted isoelectronic $^{73}$Ge in the active dot region with an energy of \SI{12}{\kilo \electronvolt} and a fluence of $4\cdot10^{13}/\text{cm}^2$, aiming for around \SI{800}{ppm} at the interface. An \SI{8}{\nano \meter} high-quality SiO$_{2}$ was thermally grown on the silicon substrate. Al gates were fabricated using an electron-beam lithography, thermal deposition of Al and lift-off process. Each Al electrode is electrically isolated by a layer of aluminum oxide of ~\SI{4}{\nano \meter} formed via thermal oxidation on a hotplate at \SI{150}{\degreeCelsius}. The devices were designed with a plunger gate width of \SI{35}{\nano \meter} and a gate pitch as small as \SI{55}{\nano \meter}. This allows a \SI{20}{\nano \meter} gap between the plunger gates for the J gates. 

\subsection{Measurement setup}\label{methods:measurement_setup}
\noindent
The device was measured in a Bluefors LD400 dilution refrigerator and mounted on the cold finger. An external d.c. magnetic field pointing in the [110] direction of the Si lattice was supplied by a superconducting split coil magnet and an American Magnetics AMI430 controller. We used Stanford SIM928 rechargeable isolated voltage sources to supply d.c. voltages through filtered lines with a bandwidth from 0 to $\SI{20}{\hertz}$. Dynamic voltage pulses were generated with a field-programmable gate array (FPGA). In this experiment, the Quantum Machines (QM) Operator-X (OPX) was combined with d.c. biases using custom linear bias combiners at room temperature. The OPX has a sampling time of $\SI{1}{\nano\second}$. The dynamic pulse lines in the dilution refrigerator have a bandwidth of 0 to $\SI{50}{\mega\hertz}$, which translates to a minimum rise time of $\SI{20}{\nano\second}$. Microwave pulses were generated by a Keysight PSG8267D Vector Signal Generator, with in-phase and quadrature (I/Q) and pulse modulation waveforms generated by the QM OPX. The modulated signal spans from $\SI{250}{\kilo\hertz}$ to $\SI{44}{\giga\hertz}$, but is bandwidth-limited by the cryostat line and a DC block.

The charge sensor comprises a single-island SET connected to a tank circuit for reflectometry measurement. The return signal was amplified by a Cosmic Microwave Technology CITFL1 LNA at the $\SI{4}{\kelvin}$ stage followed by one ZX60-P103LN+ and one Mini-circuits ZFL-1000LN+ LNAs at room temperature. All three amplifiers were powered by a Keysight E36312A Triple Output Programmable DC Power Supply. The Quantum Machines OPX generated the tones for the RF-SET, and digitized as well as demodulated the signals after the amplification.

\section{Data Availability}
\noindent The data supporting this work are available in a \href{https://doi.org/10.5281/zenodo.17255260}{Zenodo repository}.

\section*{Acknowledgments}
\noindent This research was funded by the Australian Research Council Centre of Excellence for Quantum Computation and Communication Technology (CE170100012) and the US Army Research Office (Contracts no. W911NF-17-1-0200 and W911NF-23-1-0113). We acknowledge the facilities, and the scientific and technical assistance provided by the UNSW node of the Australian National Fabrication Facility (ANFF), and the Heavy Ion Accelerators (HIA) node at the Australian National University. ANFF and HIA are supported by the Australian Government through the National Collaborative Research Infrastructure Strategy (NCRIS) program. 
P. S. acknowledges support from the Sydney Quantum Academy and the Baxter Charitable Foundation.

% \section{Author information}
\subsection{Author Contributions}\noindent
V. M. and A. L. devised the project.
W. H. L., K. W. C., and F. E. H. designed and fabricated the device. 
S. Q. L., J. C. M., and B. C. J. prepared and performed the isotope-selective ion implantation. 
P. S. and G. G. conducted the experiments with A. L.'s supervision and input from T. T., T. B., R. Y. S., A. S., A. M., C. H. Y., C. C. E. and A. S. D..
S. S. and T. B. assisted with the experimental setup.
G. G. performed initial cryogenic device screening characterization with W. H. L.'s supervision.
J. D. C. performed the electron wavefunction simulations.
P. S., G. G., and A. L. wrote the manuscript, with input from all authors.

\section{Competing Interests}\noindent
A. S. D. is CEO and a director of Diraq Pty Ltd. T. T., W. H. L., S. S., T. B., J. D. C., F. E. H., K. W. C., C. C. E., A. S., C. H. Y., A. S. D., and A. L. declare equity interest in Diraq. Other authors declare no competing interest.

%\section{Supplementary material}
% \subsection{Single qubit metrics}
%\section*{Extended data}
\setcounter{figure}{0}
\setcounter{table}{0}
\captionsetup[figure]{name={\bf{Extended Data Fig.}},justification=justified,font=small,singlelinecheck=false}

\end{document}